# Extremely strong coupling s-wave superconductivity in the medium-entropy alloy TiHfNbTa

Lingyong Zeng[1,#], Xunwu Hu[2,#], Mebrouka Boubeche[3,#], Kuan Li[1], Longfu Li[1], Peifei Yu[1], Kangwang Wang[1], Chao Zhang[1], Kui Jin[3,4,5,], Dao-Xin Yao[2,*], Huixia Luo[1,*]

[1]School of Materials Science and Engineering, State Key Laboratory of Optoelectronic Materials and Technologies, Key Lab of Polymer Composite & Functional Materials, Guangdong Provincial Key Laboratory of Magnetoelectric Physics and Devices, Sun Yat-Sen University, No. 135, Xingang Xi Road, Guangzhou, 510275, P. R. China

[2]Guangdong Provincial Key Laboratory of Magnetoelectric Physics and Devices, Center for Neutron Science and Technology, School of Physics, Sun Yat-Sen University, Guangzhou, 510275, China

[3]Songshan Lake Materials Laboratory, Building A1, University Innovation Town, Dongguan City, Guang Dong Province, 523808 China

[4]Beijing National Laboratory for Condensed Matter Physics, Institute of Physics, Chinese Academy of Sciences, Beijing, China

[5]Key Laboratory of Vacuum Physics, School of Physical Sciences, University of Chinese Academy of Sciences, Beijing, China

**Abstract:** Here we report a TiHfNbTa bulk medium-entropy alloy (MEA) superconductor crystallized in the body-centered cubic structure with the unit cell parameter a = 3.35925 Å, which is synthesized by an arc melting method. Superconducting properties of the TiHfNbTa are studied by employing magnetic susceptibility, resistivity, and specific heat measurements. Experimental results show a bulk superconducting transition temperature ($T_c$) of around 6.75 K. The lower and upper critical fields for TiHfNbTa are 45.8 mT and 10.46 T, respectively. First-principles calculations show that the *d* electron of Ti, Hf, Nb, and Ta is the main contribution near the Fermi level. Our results indicate that the superconductivity is a conventional s-wave type with extremely strong coupling ($\Delta C_{el}/\gamma_n T_c = 2.88$, $2\Delta_0/k_B T_c = 5.02$, and $\lambda_{ep} = 2.77$). The extremely strong coupling behavior in the s-wave type TiHfNbTa MEA superconductor is unusual because it generally happens in cuprates, pnictides, and other unconventional superconductors.

# 1 Introduction

High- and medium-entropy alloys (HEAs & MEAs), since their advent [1,2], have aroused increasing interest [3-5]. HEAs & MEAs are generally composed of a variety of metallic elements, opening up new avenues for designing novel functional materials. HEAs are alloys that contain between 5 % and 35 % of each of five or more metallic elements and have an idealized entropy of mixing with amplitudes of -1.6 R (R is the gas constant) or larger [3-5], on the other hand, MEAs have a mixed entropy between 1 R and 1.5 R [6,7]. HEAs & MEAs mainly crystallize in simple structures, such as hexagonal-closed packed (HCP) or body- or face-centered cubic (BCC, FCC) lattices with mixed site occupancy [3-6]. The medium or high mixing entropy significantly affects the free energy, stabilizing the MEAs or HEAs phases at ambient temperature. These alloys may offer avenues for excellent mechanical properties and high environmental stability, including high hardness [8], high thermal stability [9,10], and improved resistance to oxidation and corrosion [11]. Due to the slightly different sizes of the atoms in the elements, the high mechanical hardness and novel physical properties of MEAs & HEAs are considered to result from their large atomic disorder.

From the point of view of condensed matter physics, one of the most fascinating properties discovered in MEAs and HEAs is superconductivity (SC). In 2014, the first HEA superconductor was reported in the Ti-Zr-Hf-Nb-Ta system [12]. There have been more and more reports of HEA superconductors since then. The currently known HEA superconductors have been reported in BCC-, CsCl-, HCP-, and α-Mn-type structures [6,13]. Moreover, the critical temperature ($T_c$) of HEA superconductors is robust against high pressure or disorder [14,15] and displays a different dependence on the valence electron count (VEC) from those of both amorphous and crystalline alloy superconductors [15-17]. The HEA superconductor synthesized by a spark plasma sintering method shows stronger vortex pinning and significant critical current density than the arc-melted HEA sample [18,19]. However, the SC of quaternary MEAs is less reported than that of HEAs and conventional binary alloys. The MoReRu-based MEA superconductors have been reported and the optimal valence electron concentration of this HCP-type MEA system is near 7.0 [20]. And the structural transformation was first reported by adding carbon into the hexagonal MoReRu MEA [21]. Besides, the Tc of quaternary TiZrHfNb MEA can effectively increase by applying external pressure [22]. In the design of MEA and HEA, the Nb element with relatively high Tc is one of the most used elements. And among the alloy superconductors, NbTi alloy is a well-known commercial superconducting material with superior properties, such as accessible superconducting transition temperature, high critical current density, high critical magnetic field, easy processability, etc [23-25]. Furthermore, the NbTi alloy is the most robust among the known superconductors under pressure [26]. The $T_c$ of NbTi alloy is raised from ~ 9.6 K to ~ 19.1 K under high pressure.

Inspired by all these discoveries, we set out to find a new MEA superconductor with unique properties. Here we report on the formation, crystal structure, and physical properties of a quaternary MEA TiHfNbTa, which is nearly optimally counting the valence electrons 4.6 e/a. The TiHfNbTa MEA shows bulk type-II SC with Tc of about 6.75 K, which is determined by resistivity, magnetic susceptibility, and specific heat measurements. The results show that the SC is of a conventional s-wave type and lies in the extremely strong coupling regime ($\Delta C_{el}/\gamma_n T_c = 2.88$, $2\Delta_0/k_B T_c = 5.02$, and $\lambda_{ep} = 2.77$).

**2 Materials and method**

A polycrystalline TiHfNbTa MEA was prepared through a conventional arc-melting method, where the stoichiometric amounts of metal elements (> 99.9% purity) in an argon atmosphere with the water-cooled hearth. We re-melted several times with the inversion of the button to ensure homogeneous mixing of the metal elements. The crystal structure and phase purity were determined using X-ray diffraction (XRD). Since the obtained samples are too hard to grind, we characterized the flat surface of the samples after being polished with sandpapers. The XRD data were collected from 10 ° to 100 ° with a constant scan speed of 1 o/min at room temperature using the MiniFlex of Rigaku (Cu Kα1 radiation). The scanning electron microscope (SEM), energy dispersive X-ray spectroscopy (EDX), and backscattered electron micrograph

(BSEM) were used to determine the chemical compositions. The magnetization, resistivity, and heat capacity measurements were performed in a physical property measurement system (PPMS, Quantum Design).

The 2 × 2 × 2 supercell with 16 atoms is built by the "mcsqs" code of the Alloy Theoretic Automated Toolkit (ATAT) [27] to model the chemically disordered solutions and find the best sqs structure that most satisfied the correction function for alloy TiHfNbTa. We perform first-principles calculations as implemented in the Vienna ab initio simulation package (VASP) [28,29] based on the density functional theory. The generalized gradient approximation (GGA) of Perdew-Burke-Ernzerhof (PBE) [30] is used for exchange-correlation functions. The projector augmented-wave (PAW) potential [31] with a 400 eV plane-wave cutoff energy is employed. A 6×6×6 k-points mesh according to the Monkhorst-Pack scheme is used for all the calculations. The atomic positions of the supercell are fully optimized until forces are less than 0.01 eV/Å, and the energy convergence criterion was set to be 10-6 eV for the electronic self-consistent loop.

## 3 Discussion and conclusions

Fig. 1 displays the bulk XRD pattern for the MEA TiHfNbTa sample. It is clear from the XRD pattern that the product is pure TiHfNbTa with no remaining impurities. The mild broadening of the peak signal was due to the high degree of the disorder. All observed XRD peaks are labeled with their respective Miller indices. The XRD data were easily indexed within the BCC structure (space group Im-3m, No. 229) with a lattice parameter of a = 3.35925 Å. From the SEM and BSEM images, and elemental mappings by EDX in Fig. 1(c), we can see that the studied TiHfNbTa compound is microscopically homogeneous. The EDX results give the chemical formula of $Ti_{0.953}Hf_{1.026}Nb_{1.007}Ta_{1.014}$ (see Fig. 1(b)), which is very close to the design value. The slight deviations may be caused by the limited accuracy of EDX and the unevenness of the sample surface.

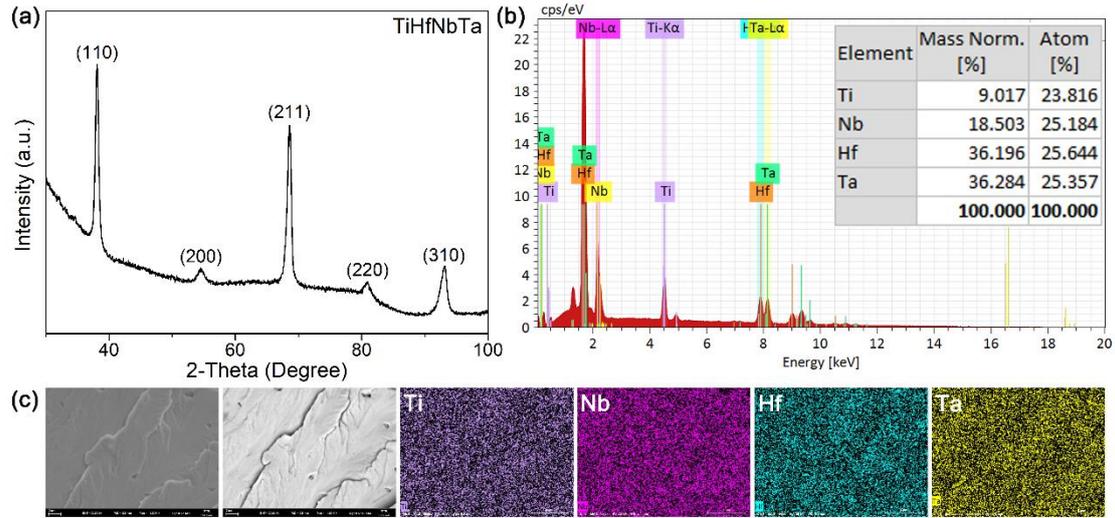

Fig. 1 (a) XRD pattern of the TiHfNbTa MEA sample. (b) EDX spectrum of the TiHfNbTa MEA, the inset displays the element ratios of the TiHfNbTa MEA. (c) SEM, BSEM images, and EDX elemental mappings for the TiHfNbTa MEA.

Fig. 2(a) shows the magnetization data (defined as $\chi_v = \dfrac{M_v}{H}$, where H is the applied magnetic field and $M_v$ is the volume magnetization) measured in the zero-field-cooled (ZFC) process under an applied field of 2 mT. A strong diamagnetic signal observed below the critical temperature $T_c$ = 6.75 K confirms the appearance of SC in the TiHfNbTa MEA samples. As explained below, the magnetization curves were further corrected by the demagnetization factor N calculated from the 1.8 K isothermal M(H) data. The $4\pi\chi_v(1-N)$ (ZFC) value approaches -1 at 1.8 K, which indicates a 100 % Meissner volume fraction. Fig. 2(b) shows the magnetic isotherms measured over a temperature range of 1.8 - 5 K. In the case of a perfect response to the field, $M_{fit}$ = m + nH is determined by conducting linear fitting for the low-filed region of the magnetization data at 1.8 K. The demagnetization factor N was then obtained from the formula $-n = \dfrac{1}{4\pi(1-N)}$. Based on the slope value (n), the N value is determined as 0.20. Fig. 2(c) shows the M - $M_{fit}$ curves at different temperatures. A black dashed line M - $M_{fit}$ = 1.40 emu/cm$^3$ is presented in Fig. 2(c), to indicate from where the data points in Fig. 2(d) have been extracted. It should be clarified that the value 1.40 emu/cm$^3$ is less than 2 % of M at 1.8 K so that the lower critical field μ₀H$_{c1}$ can be accurately calculated. The extracted points are fitted with the empirical formula $\mu_0 H_{c1}^*(T) = \mu_0 H_{c1}^*(0)(1-(T/T_c)^2)$, where Tc is superconducting transition temperature and $\mu_0 H_{c1}^*(0)$ is the lower critical field at 0 K. The $\mu_0 H_{c1}^*(0)$ is determined to be 36.6(3) mT. Furthermore, when the demagnetization factor N is taken into account, the μ₀H$_{c1}$(0) can be estimated by $\mu_0 H_{c1}(0) = \mu_0 H_{c1}^*(0)/(1-N)$. The estimated μ₀H$_{c1}$(0) = 45.8 mT for the TiHfNbTa MEA sample.

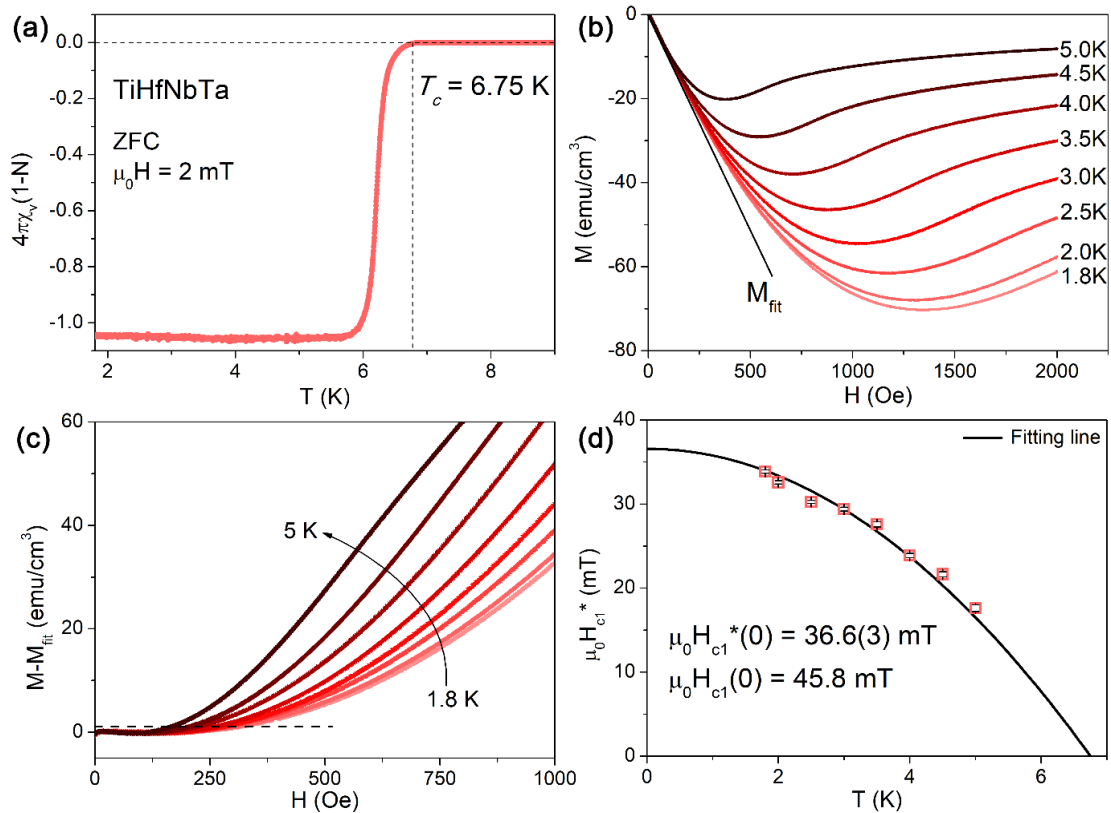

Fig. 2 (a) The temperature-dependent ZFC magnetic susceptibility for the TiHfNbTa MEA sample. (b) The isothermal magnetization curves over a temperature range of 1.8 - 5 K for TiHfNbTa MEA. (c) The difference between M and M$_{fit}$ at 0 - 1000 Oe under several temperatures. (d) The temperature dependence of the effective lower critical field for the TiHfNbTa MEA sample.

Fig. 3(a) displays the temperature-dependent electrical resistivity ρ(T) from 300

K to 1.8 K of TiHfNbTa MEA. The low-temperature resistivity under various magnetic fields ($\mu_0 H$ = 0 – 9 T) is emphasized in the inset of Fig. 3(a). At low temperatures, SC manifests as a sharp drop in $\rho(T)$, down to $\rho = 0$ value. The transition temperature from the $\rho(T)$ data is $T_c$ = 6.77 K, where Tc is defined as a 50 % decrease in resistivity, concerning the normal state value. When the magnetic field increases, Tc shifts to a lower temperature, and the superconducting transition becomes broader. We obtained the temperature variation of the upper critical field ($\mu_0 H_{c2}$) for the TiHfNbTa MEA sample with the same criterion as for zero-field resistivity data, displayed in Fig. 3(b). The resulting slope ($d\mu_0 H_{c2}/dT$) is -1.8418 T/K for the TiHfNbTa MEA compound. The data were fitted with the Werthamer-Helfand-Hohenberg (WHH) formula: . Using Tc = 6.77 K, the dirty limit $\mu_0 H_{c2}(0)_{WHH}$ value was calculated to be 8.64 T.

To explore the SC nature of the TiHfNbTa MEA, we also conducted systematic magnetotransport measurements under a perpendicular field at different temperatures (see Fig. 3(c)). We define the value of resistivity down to 50 % of $\rho_N$ as $\mu_0 H_{c2}$. The phase diagram of $\mu_0 H_{c2}$ versus T is therefore displayed in Fig. 3(d). The data were fitted with the Ginzburg-Landan (GL) formula: The GL model fits the experimental data satisfactorily in the entire temperature range. The $\mu_0 H_{c2}(0)_{GL}$ is estimated to be 10.46(3) T, which is lower than that of NbTi alloy (15.4 T) [26]. Based on BCS theory, the Pauli limiting field for a superconductor can be described by $\mu_0 H^P = 1.85*T_c$, that is, 12.52 T for the TiHfNbTa MEA sample, which is larger than the upper critical field.

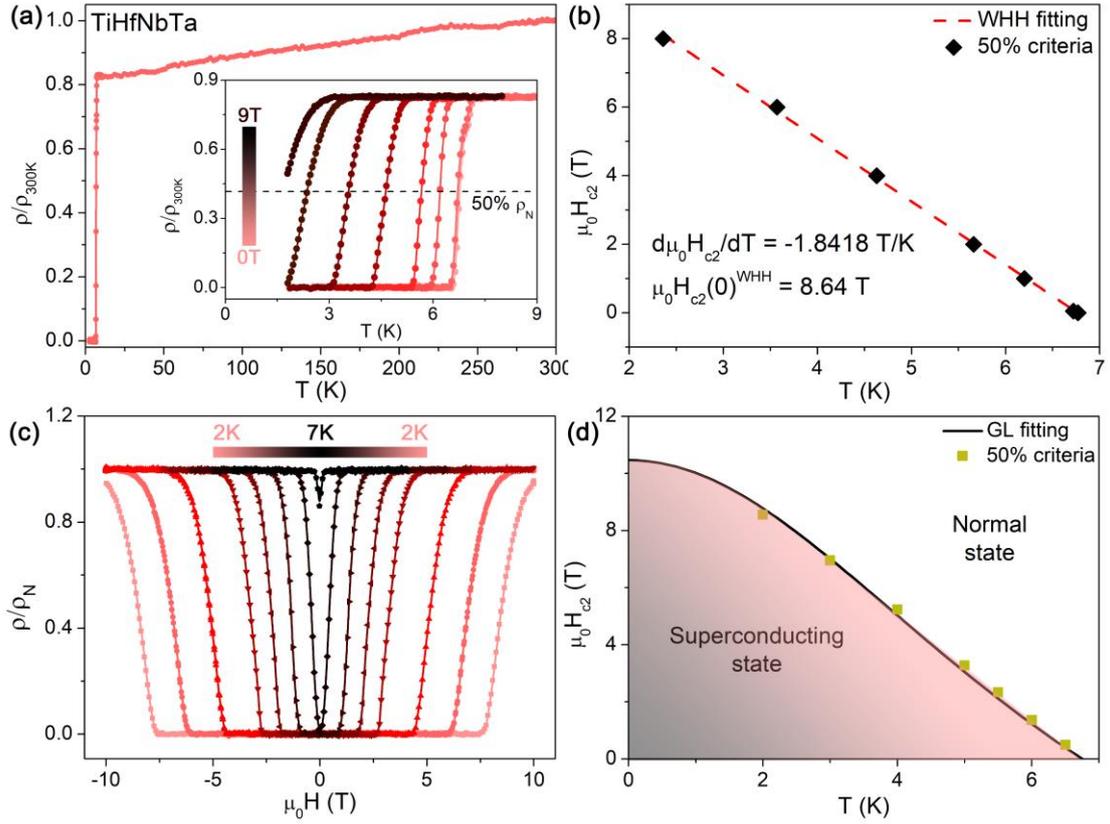

Fig. 3 (a) The temperature-dependent resistivity of the TiHfNbTa MEA sample. The inset displays the resistivity data from 1.8 to 9 K under different magnetic fields from 0 to 9 T. (b) The temperature dependence of the $\mu_0H_{c2}$ and the WHH model fitting. (c) Perpendicular field magnetoresistance at several temperatures. (d) The temperature dependence of the $\mu_0H_{c2}$ fitting with the GL model.

Based on the results of $\mu_0H_{c1}(0)$ = 45.8 mT and $\mu_0H_{c2}(0)$ = 10.46 T, a range of superconducting parameters can be calculated and extracted. The GL coherence length ($\xi_{GL}(0)$) of TiHfNbTa is estimated as 56 Å by the relation: $\xi_{GL}^2(0) = \dfrac{\phi_0}{2\pi H_{c2}(0)}$ , in which the $\Phi_0$ = h/2e stands for the magnetic flux quantum. The $\xi_{GL}(0)$ value of TiHfNbTa MEA is smaller than that in Kagome lattice superconductors (e.g., LaIr$_3$Ga$_2$ (84 Å), LaRu$_3$Si$_2$ (107 Å)) [32,33] and heavy fermion superconductors (e.g., UPd$_2$Al$_3$ (85 Å), CeIrIn$_5$ (71 Å), CeRhIn$_5$ (57 Å)) [34,35]. It suggests that stronger electron-electron interaction exists in the TiHfNbTa MEA than is reported for 132-type Kagome lattice superconductors and near the heavy fermion superconductors. Furthermore, the superconducting penetration depth at 0 K ($\lambda_{GL}(0)$) can be obtained from $\mu_0 H_{c1}(0) = \frac{\phi_0}{4\pi\lambda_{GL}^2(0)} \ln \frac{\lambda_{GL}(0)}{\xi_{GL}(0)}$, to give $\lambda_{GL}(0)$ = 1022 Å. Consequently, from the relation $K_{GL}(0) = \frac{\lambda_{GL}(0)}{\xi_{GL}(0)}$, we find the GL parameter $K_{GL}(0)$ = 18.25. This value is larger than $1/\sqrt{2}$, indicating the TiHfNbTa MEA is a type-II superconductor. The thermodynamic critical field ($\mu_0 H_c(0)$) is calculated from $\mu_0 H_{c1}(0) \times \mu_0 H_{c2}(0) = \mu_0 H_c^2(0) \times \ln K_{GL}(0)$ to be 0.41 T. Table 1 summarizes all these superconducting parameters.

To take a further insight into the SC, as well as the thermodynamic properties of TiHfNbTa MEA, we measured the specific heat ($C_p$) under a zero applied field. Fig. 4(a) shows the $C_p/T(T^2)$ curves at lower temperatures. Bulk SC is evident by a sharp jump at $T_c$ = 6.30 K. The normal state data were fitted with the equation $C_p/T = \gamma_n + \beta T^2$, where β is the specific heat coefficient of the lattice part, and γn represents the Sommerfeld constant of the normal state. The fit gives *β* = 0.242(8) mJ/mol/K$^4$ and γn = 4.698(5) mJ/mol/K$^2$. The Debye temperature is given by $\Theta_D = (12\pi 4nR/5\beta)^{1/3}$, where R represents a gas constant, and n represents the number of atoms in the formula unit. We can obtain $\Theta_D$ = 199.9 K for the TiHfNbTa MEA.

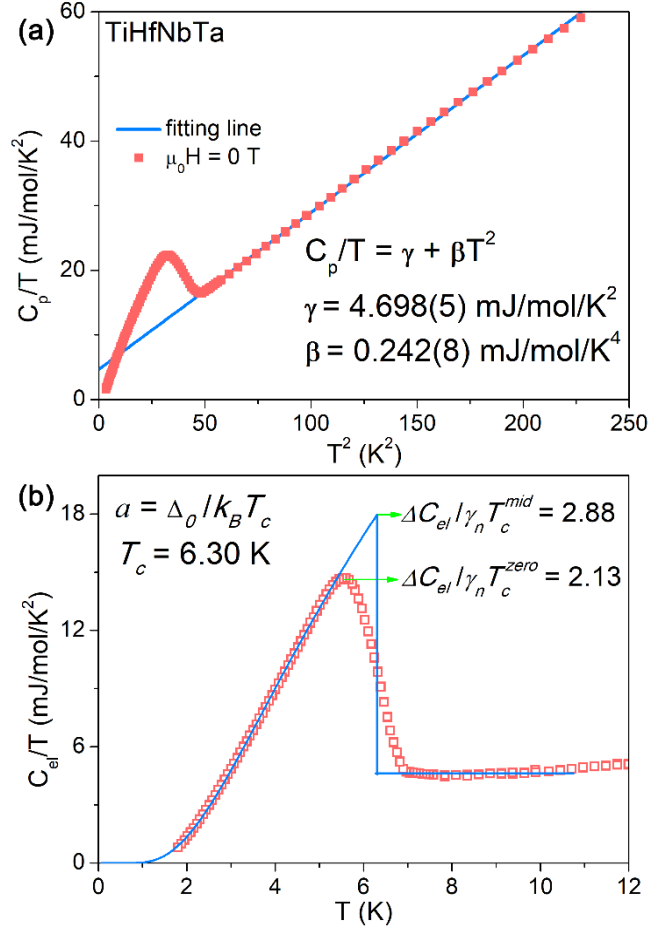

Fig. 4 (a) The $C_p/T$ vs $T^2$ curves over a temperature range of 1.8 - 15 K, fitted with low-temperature Debye model $C_p/T = \gamma^n + \beta T^2$. (b) The specific heat data over a temperature range of 1.8 - 12 K fitted with the α-model.

By subtracting the phonon terms from $C_p$, we can obtain the electronic contribution to $C_p$. Fig. 4(b) displays the temperature-dependent electronic specific heat ($C_{el}$). Note that the normalized $C_{el}$ jump at $T_c^{mid}$ ($\Delta C_{el}/\gamma_n T_c^{mid}$) and $T_c^{zero}$ ($\Delta C_{el}/\gamma_n T_c^{zero}$) for the TiHfNbTa MEA are 2.88 and 2.13, respectively. It is well known that the ratio $\Delta C_{el}/\gamma_n T_c$ is used to measure the strength of the electron coupling. Both values of $\Delta C_{el}/\gamma_n T_c$ are much larger than the Bardeen-Cooper-Schrieffer (BCS) weak coupling ratio (1.43), indicating strong coupling in the TiHfNbTa MEA.

The electron-phonon coupling constant ($\lambda_{ep}$) is estimated by using the inverted McMillan equation. The μ* represents the Coulomb pseudopotential parameter and has the typical value of 0.13 [36-40]. The value of $\lambda_{ep}$ is determined to be 0.83, which is close to the value of 0.9 for another strongly coupled superconductor $Ta_{1/6}Nb_{2/6}Hf_{1/6}Zr_{1/6}Ti_{1/6}$ [18]. The ratio $\Delta C_{el}/\gamma_n T_c$ has demonstrated that TiHfNbTa MEA is a strongly coupled superconductor. Therefore, the electron-phonon coupling parameter ($\lambda_{ep}$) for a strong coupling superconductor is further determined using the McMillan equation modified by Allen and Dynes [41,42]:

$$T_c = \frac{\omega_{\ln}}{1.2} \exp[\frac{1.04(1+\lambda_{ep})}{\mu^*(1+0.62\lambda_{ep})-\lambda_{ep}}],$$ where the logarithmic average phonon frequency $\omega\_\ln$ is written as: . We can obtain $\omega_{\ln}$ = 38 K and $\lambda_{ep}$ = 2.77. This value will be used as the experimental estimation of $\lambda_{ep}$ in Fig. 5. The large value of $\lambda_{ep}$ confirms the strong coupling nature of the superconducting pairing. Based on the $\lambda_{ep}$ and $\gamma_n$, we can calculate the electronic states at the Fermi energy $N(E_F)$ from the formula

$$N(E_F) = \frac{3\gamma_n}{\pi^2 k_B^2 (1+\lambda_{ep})}.$$ $N(E_F)$ was calculated to be 2.49 eV$^{-1}$ f.u.$^{-1}$ of TiHfNbTa MEA.

Moreover, it can be seen that a conventional s-wave gap function can well fit the data, and the so-called α-model is applied, indicating a fully gapped material [43]. In the α-model, the angular independent gap function $\Delta(T)$ can be written as $\Delta(T) = \alpha/\alpha_{BCS}\Delta_{BCS}(T)$ where $\alpha_{BCS}$ = 1.76 represents the weak coupling gap ratio. Fittings to the $C_{el}$ data are displayed in Fig. 4(b), from which we get the superconducting gap value at zero temperature $\Delta_0$ = 1.36 meV. The coupling strength $2\Delta_0/k_B T_c$ is thus estimated to be 5.02. Again, this value exceeds the weak coupling BCS value of 3.52, evidencing strong coupled SC in the TiHfNbTa MEA. The strong coupling s-wave SC with $2\Delta_0/k_B T_c$ = 5.02 is rather rare, only a few examples are known to date and include the Pd-Bi alloy [42], pyrochlore osmates [44], and antiperovskite phosphide superconductors [45].

Three important quantities for characterizing the SC are $\Delta C_{el}/\gamma_n T_c$, $2\Delta_0/k_B T_c$, and $\lambda_{ep}$. The obtained value of $\Delta C_{el}/\gamma_n T_c$, $2\Delta_0/k_B T_c$, and $\lambda_{ep}$ from the analysis of specific data by the α-model and the temperature-dependent $C_{el}$ is 2.88, 5.02, and 2.77, respectively. Fig. 5 shows these important quantities characterizing the strong coupling SC as a function of the average phonon frequency ωln normalized by $T_c$. Three values are compared in Fig. 5 with those for typical strong couplings superconductors such as Pb and its alloys [42], A15 compounds [42], Chevrel phase compounds [46], β-pyrochlore oxide $KOs_2O_6$ [44], and nickel-based $LaO_{0.9}F_{0.1}NiAs$ [47]. Each symbol represents a superconductor: triangle corresponds to those studied by Carbotte [42], circle to some Chevrel phase samples, squares to $KOs_2O_6$, Inverted triangle to $LaO_{0.9}F_{0.1}NiAs$, and pentagram to TiHfNbTa MEA. Green lines serve as eye guides and the red dashed line represents the weak coupling limit BCS value. The $\Delta C_{el}/\gamma_n T_c$ and $2\Delta_0/k_B T_c$ parameters of TiHfNbTa are exactly on the green line. Among these superconductors, the TiHfNbTa MEA possesses almost the largest value for $\Delta C_{el}/\gamma_n T_c$, $2\Delta_0/k_B T_c$, and $\lambda_{ep}$, demonstrating that it is categorized as an extremely strong coupling superconductor.

In McMillan's formalism [48], the electron-phonon coupling strength can be written as $\lambda_{ep} = \left[ N(E_F)\langle I^2 \rangle \right] / \left[ M \langle \omega^2 \rangle \right]$, where $\langle \omega^2 \rangle$ is the squared phonon frequencies and $\langle I^2 \rangle$ is the squared electronic matrix element on the Fermi surface. Thus, λep can be enhanced by decreasing the $\langle \omega^2 \rangle$ (soften the lattice) and increasing the $N(E_F)$. Comparing the Nb element, NbTi alloy, and TiHfNbTa MEA with the same BCC structure (see Table 1). The $N(E_F)$ of TiHfNbTa (2.49 $eV^{-1}$ f.u.$^{-1}$) is much larger than that of Nb (~1 $eV^{-1}$ f.u.$^{-1}$) and close to that of NbTi alloy (~2.50 $eV^{-1}$ f.u.$^{-1}$) [49]. And the lattice of TiHfNbTa is most softened, as evidenced by the smallest of $\Theta_D$. These two factors can make TiHfNbTa MEA have stronger electron-phonon coupling.

Table 1 summarizes the relevant superconducting parameters of Nb elements and Nb-based alloys [18,49,50]. Although the Tc of TiHfNbTa is not too high, TiHfNbTa MEA has the strongest electron-phonon coupling strength in the Nb-based alloys. The strongly coupled behavior is also observed in $Ta_{1/6}Nb_{2/6}Hf_{1/6}Zr_{1/6}Ti_{1/6}$ HEA [18]. And we propose that the low-lying phonons might be key ingredients in realizing strong coupling SC of TiHfNbTa MEA. Evidence of strong coupling caused by low-lying phonons may be captured in normal-state resistivity data. ρ(T) shows almost T-linear resistivity below 40 K (see Fig. S1), implying that there present low-energy (~ meV) modes responsible for scatterings [45]. And these low-energy modes can be attributed to low-lying phonons [45]. Many strongly coupled superconductors with large λep values have a variety of low-lying phonons. The Chevrel phase compounds are considered "molecular crystals" with a low-lying phonon mode [46]. And the A15 compounds display soft phonon modes associated with a cubic-to-tetragonal transition [51].

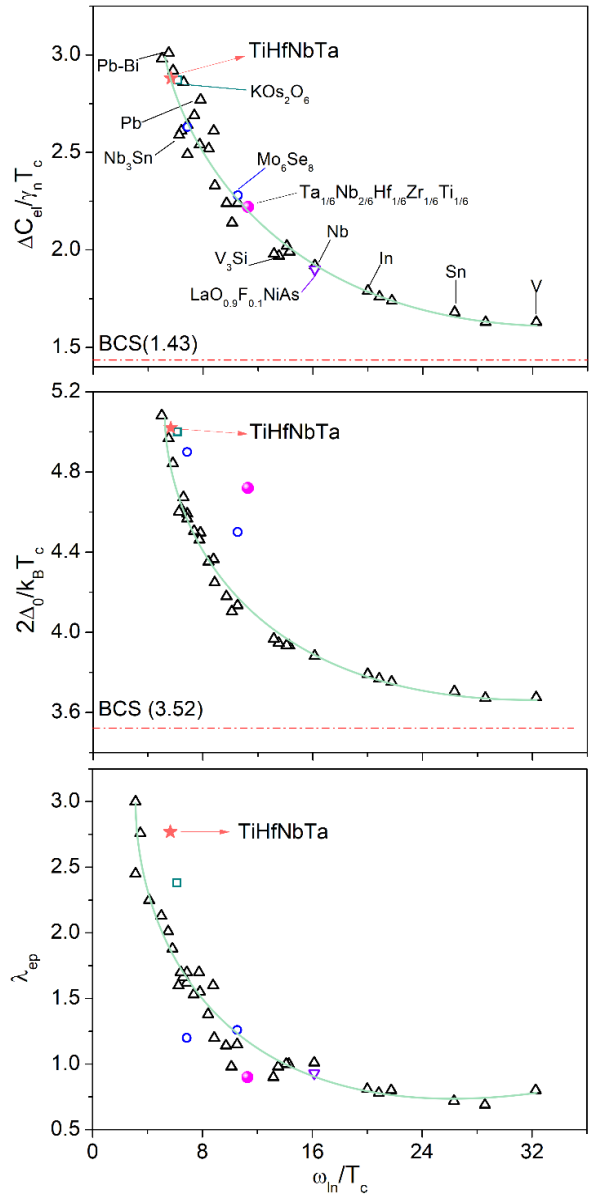

Fig. 5 Three important quantities characterizing the strong coupling SC as a function of the average phonon frequency ωln normalized by $T_c$. Every symbol is a superconductor (ref. 18,39,41,43,44).

Table 1. The relevant superconducting parameters of Nb element and typical Nb-based alloys.

| | TiHfNbTa | Nb [49] | NbTi [49] | Nb-Ta-V [50] | $Ta_{1/6}Nb_{2/6}Hf_{1/6}Zr_{1/6}Ti_{1/6}$ [18] | BCS |
|---|---|---|---|---|---|---|
| $T_c$ (K) | 6.75 | 9.23 | 9.71 | ~4-5 | 7.85 | |
| $\mu_0H_{c1}$ (mT) | 45.8 | | | | 23 | |
| $\mu_0H_{c2}$ (T) | 10.46 | 0.41 | ~14 | | 12.05 | |
| $\omega_{ln}$ (K) | 38 | | | | | |
| $\gamma_n$ (mJ/mol/K$^2$) | 4.70 | 7.8 | 10.76 | ~6-8 | 7.45 | |
| $\Delta C_{el}/\gamma_n T_c$ | 2.88 | 2.07 | ~1.85 | | 2.22 | 1.43 |
| $2\Delta_0/k_BT_c$ | 5.02 | 3.8 | ~2.3 | | 4.72 | 3.53 |
| $\Theta_D$ | 199.9 | 276 | 236 | | 192.3 | |
| $\lambda_{ep}{}^a$ | 0.83 | 0.82 | 1.67 | ~0.8-1 | 0.9 | |

As shown in Fig. 6, the local density of states (DOS) and the partial DOS of s, p, and d orbitals of Ti, Hf, Nb, and Ta, and the total density of states (TDOS) for alloy TiHfNbTa was calculated using first principles. We also consider 24 assumed random atomic arrangements and structural configurations to investigate the influence of the disorder on the electronic properties of the alloy TiHfNbTa. The overall shape of the averaged (24 randoms configurations) and sqs (structure constructed using the mcsqs code (see Fig. S2)) TDOS are almost identical, which means that the disorder of the structure has a tiny effect on the TDOS of alloy TiHfNbTa. Therefore, we only consider the sqs structure in the rest of the calculations. The lattice parameter fitted by the Birch–Murnaghan equation of state is 3.35957 Å, which is in excellent agreement with the experimental lattice parameter (See Fig. S3). As can be seen in Fig. 6(a), the TDOS passing through the Fermi level suggests its metallic properties (See Fig. S4). The density of states is about 1.5 states eV$^{-1}$f.u.$^{-1}$ at the Fermi level. Inserting this value into the relationship $\frac{1}{3}\pi k_B N_A N(E_F)$, the theoretical Sommerfeld coefficient γth is estimated to be 3.54 mJ/mol/K$^2$, which is close to the experimental value $\gamma_n$ = 4.698(5) mJ/mol/K$^2$. The local DOS diagram shows that the Ti and Hf atoms are the largest contribution to TDOS near the Fermi level, and Ta and Nb atoms also have some contribution. As shown in Fig 6(b-e), the projected DOS with angular momentum reveal that the d-electron of all elements are the main contributions to the TDOS, i.e., 3d for Ti, 4d for Nb, and 5d for Ta and Hf, although the p-electron has some contribution.

These results indicate that the superconductivity may mainly originate from the d-electrons of Ti, Hf, Nb, and Ta, which can cause a strong coupling in the material.

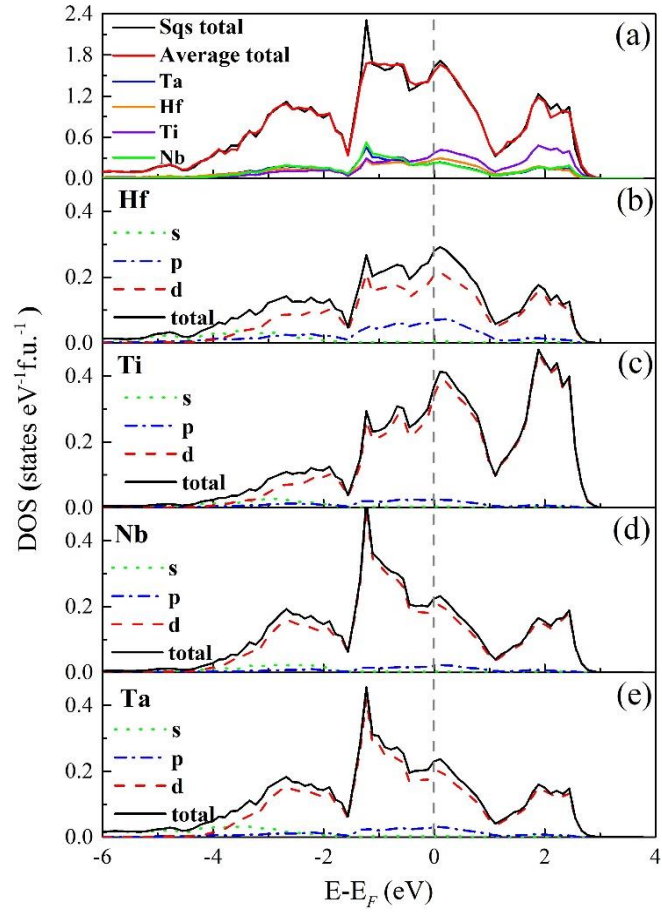

Fig. 6 (a)Total (black solid line) and local DOS of each element calculated for alloy TiHfNbTa with the structure built by mcsqs code. Average total DOS (red solid line) for 24 assumed random atomic arrangements. (b)-(e) Projected DOS with angular momentum decomposition of each element. The gray dashed line indicates the Fermi level.

**Conclusion**

We synthesized the TiHfNbTa MEA by the arc-melting method as a single phase with BCC crystal structure (Im-3m). The superconducting phase transition is observed at $T_c$ ~ 6.75 K and confirmed that it is a type-II superconductor showing relatively high lower and upper critical fields of 45.8 mT and 10.46 T, respectively. First-principles calculations show that the 3/4/5d electron of Ti, Hf, Nb, and Ta is the main contribution near the Fermi level. Moreover, TiHfNbTa MEA is a fully gapped superconductor with extremely strong electron-phonon coupling, as evidenced by the large values of $\Delta C_{el}/\gamma_n T_c = 2.88$, $2\Delta 0/k_B T_c = 5.02$, and $\lambda_{ep} = 2.77$.

## Acknowledgments

This work is supported by the National Natural Science Foundation of China (12274471, 11922415), Guangdong Basic and Applied Basic Research Foundation (2022A1515011168, 2019A1515011718), the Key Research & Development Program of Guangdong Province, China (2019B110209003). Mebrouka Boubeche is supported by the Foreign Young Talents Program of China (22KW041C211). K. Jin is supported by the Key-Area Research and Development Program of Guangdong Province (Grant No. 2020B0101340002). D. X. Yao and X. Hu are supported by NKRDPC-2022YFA1402802, NKRDPC-2018YFA0306001, NSFC-11974432, NSFC-92165204, Leading Talent Program of Guangdong Special Projects (201626003), and Shenzhen International Quantum Academy (Grant No. SIQA202102).

**References:**

[1] J. W. Yeh, S. K. Chen, S. J. Lin, J. Y. Gan, T. S. Chin, T. T. Shun, C. H. Tsau, and S. Y. Chang, Adv. Eng. Mater. 6, 299 (2004).

[2] B. Cantor, I. T. H. Chang, P. Knight, and A. J. B. Vincent, Mater. Sci. Eng. A 375-377, 213 (2004).

[3] E. P. George, D. Raabe, and R. O. Ritchie, Nat. Rev. Mater. 4, 515 (2019).

[4] Y. F. Ye, Q. Wang, J. Lu, C. T. Liu, and Y. Yang, Mater. Today 19, 349 (2016).

[5] D. B. Miracle, and O. N. Senkov, Acta Mater. 122, 448 (2017).

[6] L. Sun, and R. J. Cava, Phys. Rev. Mater. 3, 090301 (2019).

[7] J. W. Yeh, JOM 65, 1759 (2013).

[8] P. Edalati, A. Mohammadi, M. Ketabchi, and K. Edalati, J. Alloys Compd. 884, 161101 (2021).

[9] M. Vaidya, K. Guruvidyathri, and B. S. Murty, J. Alloys Compd. 774, 856 (2019).

[10] T. E. Whitfield, E. J. Pickering, L. R. Owen, O. N. Senkov, D. B. Miracle, H. J. Stone, and N. G. Jones, J. Alloys Compd. 857, 157583 (2021).

[11] N. Hua, W. Wang, Q. Wang, Y. Ye, S. Lin, L. Zhang, Q. Guo, J. Brechtl, and P. K. Liaw, J. Alloys Compd. 861, 157997 (2021).

[12] P. Koželj, S. Vrtnik, A. Jelen, S. Jazbec, Z. Jagličić, S. Maiti, M. Feuerbacher, W. Steurer, and J. Dolinšek, Phys. Rev. Lett. 113, 107001 (2014).

[13] J. Kitagawa, S. Hamamoto, and N. Ishizu, Metals 10, 1078 (2020).

[14] J. Guo, H. H. Wang, F. von Rohr, Z. Wang, Y. Z. Zhou, K. Yang. A. G. Li, Q. Wu, R. J. Cava, L. and L. Sun, Proc. Natl. Acad. Sci. U.S.A. 114, 13144 (2017).

[15] F. von Rohr, M. J. Winiarski, J. Tao, T. Klimczuk, and R. J. Cava, Proc. Natl. Acad. Sci. U.S.A. 113, E7144 (2016).

[16] B. Liu, J. Wu, Y. Cui, Q. Zhu, G. Xiao, S. Wu, G. h. Cao, and Z. Ren, J. Mater. Sci. Technol. 85, 11 (2021).

[17] G. Xiao, Q. Zhu, W. Yang, Y. Cui, S. Song, G. H. Cao, and Z. Ren, Sci. China Mater. 66, 257 (2022).

[18] G. Kim, M. H. Lee, J. H. Yun, P. Rawat, S. G. Jung, W. Choi, T. S. You, S. J. Kim, and J. S. Rhyee, Acta Mater. 186, 250 (2020).

[19] J. H. Kim, R. Hidayati, S. G. Jung, Y. A. Salawu, H. J. Kim, J. H. Yun, and J. S. Rhyee, Acta Mater. 232, 117971 (2022).

[20] Y. S. Lee, and R. J. Cava, Physica C Supercond. 566, 1353520 (2019).

[21] Q. Zhu, G. Xiao, Y. Cui, W. Yang, S. Wu, G. H. Cao, and Z. Ren, Scripta Materialia

210, 114464 (2022).

[22] S. A. Uporov, R. E. Ryltsev, V. A. Sidorov, S. K. Estemirova, E. V. Sterkhov, I. A. Balyakin, and N. M. Chtchelkatchev, Intermetallics 140, 107394 (2022).

[23] M. Parizh, Y. Lvovsky, and M. Sumption, Supercond. Sci. Technol. 30, 014007 (2017).

[24] J.-H. Liu, J.-S. Cheng, and Q.-L. Wang, IEEE Trans. Appl. Supercond. 23, 4802606 (2013).

[25] J.-F. Zhang, M. Gao, K. Liu, and Z.-Y. Lu, Phys. Rev. B 102, 195140 (2020).

[26] J. Guo, G. C. Lin, S. Cai, C. Y. Xi, C. J. Zhang, W. S. Sun, Q. L. Wang, K. Yang, A. G. Li, Q. Wu, Y. H. Zhang, T. Xiang, R. J. Cava, and L. L. Sun, Adv. Mater. 31, 1807240 (2019).

[27] A. van de Walle, Calphad 33, 266 (2009).

[28] G. Kresse, and J. Hafner, Phy. Revs. B 47, 558 (1993).

[29] G. Kresse, and J. Furthmüller, Phys. Rev. B 54, 11169 (1996).

[30] J. P. Perdew, K. Burke, and M. Ernzerhof, Phy. Rev. Lett. 77, 3865 (1996).

[31] P. E. Blöchl, Phys. Rev. B 50, 17953 (1994).

[32] X. Gui, and R. J. Cava, Chem. Mater. 34, 2824 (2022).

[33] Y. Kishimoto, T. Ohno, T. Hihara, K. Sumiyama, G. Ghosh, and L. C. Gupta, J. Phys. Soc. Japan 71, 2035 (2002).

[34] C. Geibel, C. Schank, S. Thies, H. Kitazawa, C. D. Bredl, A. Böhm, M. Rau, A. Grauel, R. Caspary, R. Helfrich, U. Ahlheim, G. Weber and F. Steglich Z. Physik B - Condensed Matter 84, 1 (1991).

[35] B. D. White, J. D. Thompson, and M. B. Maple, Physica C Supercond 514, 246 (2015).

[36] L. Y. Zeng, X. W. Hu, S. Guo, G. T. Lin, J. Song, K. Li, Y. Y. He, Y. H. Huang, C.

Zhang, P. F. Yu, J. Ma, D. X. Yao, and H. X. Luo, Phys. Rev. B 106, 134501 (2022).

[37] L. Y. Zeng, X. W. Hu, N. N. Wang, J. P. Sun, P. T. Yang, M. Boubeche, S. J. Luo, Y. Y. He, J. G. Cheng, D. X. Yao, and H. X. Luo, J. Phys. Chem. Lett. 13, 2442 (2022).

[38] G. R. Xiao, Q. Q. Zhu, Y. W. Cui, W. Z. Yang, B. Z. Li, S. Q. Wu, G.-H. Cao, and Z. Ren, Sci. China Phys. Mech. Astron. 64, 107411 (2021).

[39] Z. C. Liu, B. Z. Li, Y. S. Xiao, Q. C. Duan, Y. W. Cui, Y. X. Mei, Q. Tao, S. L. Wei, S. G. Tan, Q. Jing, Q. Lu, Y. P. Sun, Y. Y. Liu, S. G. Fu, H. Jiang, Z. Ren, Z. A. Xu, C. Wang, and G. H. Cao, Sci. China Phys. Mech. Astron. 64, 277411 (2021).

[40] Y. X. Zhou, B. Li, Z. F. Lou, H. C. Chen, Q. Chen, B. J. Xu, C. X. Wu, J. H. Du, J. H. Yang, H. D. Wang, and M. H. Fang, Sci. China Phys. Mech. Astron. 64, 247411 (2021).

[41] P. B. Allen, and R. C. Dynes, Phys. Rev. B 12, 905 (1975).

[42] J. P. Carbotte, Rev. Mod. Phys. 62, 1027 (1990).

[43] D. C. Johnston, Supercond Sci Technol 26, 115011 (2013).

[44] Z. Hiroi, S. Yonezawa, Y. Nagao, and J. Yamaura, Phys. Rev. B 76, 014523 (2007).

[45] T. Takayama, K. Kuwano, D. Hirai, Y. Katsura, A. Yamamoto, and H. Takagi, Phys. Rev. Lett. 108, 237001 (2012).

[46] M. Furuyama, N. Kobayashi, and Y. Muto, Phys. Rev. B 40, 4344 (1989).

[47] Z. Li, G. F. Chen, J. Dong, Gang. Li, W. Z. Hu, D. Wu, S. K. Su, P. Zheng, T. Xiang, N. L. Wang, and J. L. Luo, Phys. Rev. B 78, 060504 (2008).

[48] W. L. McMillan, Phys. Rev. 167, 331 (1968).

[49] N. K. Sarkar, C. L. Prajapat, P. S. Ghosh, N. Garg, P. D. Babu, S. Wajhal, P. S. R. Krishna, M. R. Gonal, R. Tewari, and P. K. Mishra, Intermetallics 114, 107503 (2022).

[50] R. YU. P. Wang, and R. L. Cappelletti, J. Low Temp. Phys. 32, 481 (1978).

[51] Z. Fisk, and G. W. Webb Phys. Rev. Lett. 36, 1084 (1976).